\documentclass[twocolumn,showpacs,preprintnumbers,amsmath,amssymb,prl,aps,superscriptaddress]{revtex4} 

\usepackage{graphicx}
\usepackage{dcolumn}
\usepackage{bm}
\usepackage{color}

\begin{document}

\title{Symmetry Breaking and Enhanced Condensate Fraction\\ in a Matter-Wave Bright Soliton}
\author{Rina Kanamoto}
\affiliation{Department of Physics, Osaka City University, Osaka 558-8585, Japan}

\author{Hiroki Saito}
\affiliation{Department of Physics, Tokyo Institute of Technology, Tokyo 152-8551, Japan}
\affiliation{CREST, Japan Science and Technology Corporation (JST), Saitama 332-0012, Japan}

\author{Masahito Ueda}
\affiliation{Department of Physics, Tokyo Institute of Technology, Tokyo 152-8551, Japan}
\affiliation{CREST, Japan Science and Technology Corporation (JST), Saitama 332-0012, Japan}

\date{\today}

\begin{abstract}
An exact diagonalization study reveals that a matter-wave bright soliton 
and the Goldstone mode are simultaneously created in 
a quasi-one-dimensional attractive Bose-Einstein condensate by 
superpositions of quasi-degenerate low-lying many-body states. 
Upon formation of the soliton the maximum eigenvalue of the single-particle density matrix 
increases dramatically, indicating that a fragmented condensate 
converts into a single condensate as a consequence of the breaking of 
translation symmetry. 
\end{abstract}
\maketitle


Fragmentation of a Bose-Einstein condensate (BEC), which occurs as 
a consequence of a certain exact symmetry of the system, 
has recently been discussed in a number of articles~\cite{NS,fr,UL}. 
In contrast to the conventional BEC, characterized by 
a unique macroscopic eigenvalue in the single-particle density matrix~\cite{PO}, 
the fragmented BEC is characterized by more than one macroscopic eigenvalue~\cite{NS}. 
If the system has an exact symmetry and if the many-body theory predicts 
fragmentation of the ground state, 
the Gross-Pitaevskii (GP) mean-field theory 
does not predict a fragmented condensate but approximates it with 
a single condensate whose symmetry is spontaneously broken. 
For example, a quasi-one-dimensional (1D) BEC with attractive interaction forms 
bright solitons~\cite{BS-EX}, 
which are well described by the GP theory~\cite{BS-MF}. 
Efforts to elucidate how such symmetry-broken states emerge from 
exact many-body states 
have been made in diverse systems~\cite{symmetry}. 


In this Letter, we show that the formation of a broken-symmetry soliton 
and an enhancement of the condensate fraction 
are caused by superpositions of the low-lying states of the 
symmetry-preserving many-body Hamiltonian. 
We find that the many-body spectrum exhibits a number of 
quasi-degenerate states in the regime 
where the exact ground state is a fragmented condensate. 
Superposition of these quasi-degenerate levels simultaneously generates 
the broken-symmetry bright soliton and the Goldstone mode, 
accompanied by a significant increase in the condensate fraction. 
By introducing a small symmetry-breaking perturbation or 
by considering the action of a quantum measurement, 
we explicitly show that the fragmented condensate 
is very fragile against the soliton formation. 
Also elucidated in the language of the many-body theory 
is the mechanism underlying a partial breaking of the quantized 
circulation in the presence of a rotating drive. 



We consider a system of $N$ attractive bosons with mass $m$ on a 1D ring 
with circumference $2\pi R$. Length and energy are measured in units of $R$ 
and $\hbar^2/(2mR^2)$, respectively. 
The Hamiltonian for our system is given by 
\begin{eqnarray}\label{hamiltonian}
\hat{H}=\!\int_0^{2\pi}\!\!d\theta
\left[-\hat{\psi}^{\dagger}(\theta)\frac{\partial^2}{\partial\theta^2}\hat{\psi}(\theta)-\frac{\pi g}{2}
\hat{\psi}^{\dagger 2}(\theta){\hat{\psi}}^2(\theta)\right],
\end{eqnarray}
where $\hat{\psi}(\theta)$ is the field operator, which annihilates an atom at position $\theta$, 
and $g$ $(>0)$ denotes the strength of attractive interaction. 
According to the GP mean-field approximation for the Hamiltonian~(\ref{hamiltonian}), 
the ground state is either a uniform condensate or a broken-symmetry bright soliton, 
depending on whether the parameter $gN$ is below or above the critical value, $gN= 1$. 
In contrast, all eigenstates of the original Hamiltonian are 
translation invariant, and many-body theory predicts that 
the ground state is either a {\it single} ($gN\lesssim 1$)
or {\it fragmented} ($gN\gtrsim 1$) condensate~\cite{QPT}. 

\begin{figure}
\includegraphics[scale=0.43]{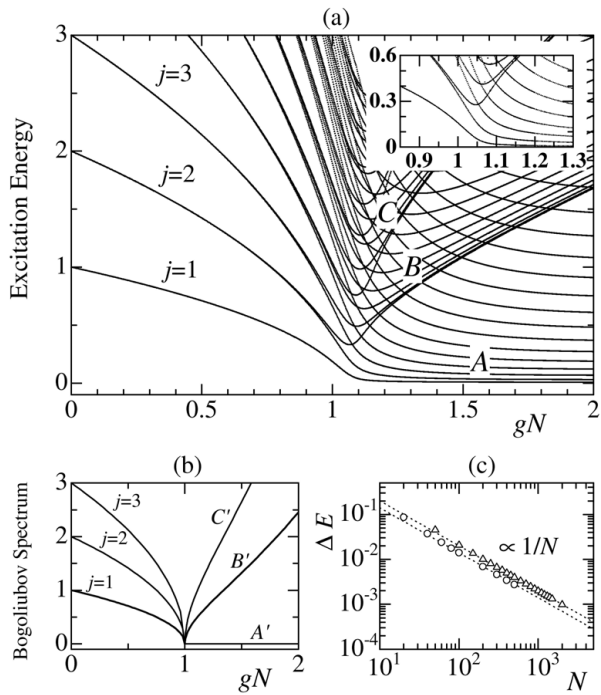}
\caption{
(a) Excitation spectrum obtained by exact diagonalization of
Hamiltonian (1) for $N=200$. 
The inset shows the corresponding result obtained with truncation 
$l_{\rm c}=2$ near the critical point.
(b) Bogoliubov spectrum corresponding to (a), where 
branch $A'$ represents the Goldstone mode, 
$B'$ the breathing mode of a bright soliton, and $C'$ the second harmonic of $B'$. 
(c) Energy gap $\Delta E$ between the ground and the first excited states 
in the many-body spectrum versus the total number of atoms $N$ with $gN=1.4$ held fixed. 
Triangles and circles denote results obtained with $l_{\rm c}=1$ and $l_{\rm c}=2$, respectively. 
}
\label{fig1}
\end{figure}


Figure~\ref{fig1} (a) shows the low-lying spectrum 
obtained by exact diagonalization of the Hamiltonian~(\ref{hamiltonian}). 
The dramatic change in the landscape of the energy spectrum around $gN \simeq 1$ 
is a consequence of the quantum phase transition between 
a single condensate and a fragmented one. 
Figure~\ref{fig1} (b) presents the Bogoliubov spectrum obtained from 
the Bogoliubov-de Gennes equations. 
By comparing Figs.~1(a) and (b), we find that 
the Bogoliubov spectrum has a one-to-one correspondence 
with the many-body spectrum for $gN \lesssim 1$. 
For $gN\gtrsim 1$, however, the many-body spectrum becomes 
much more intricate than the Bogoliubov one. 
In the Bogoliubov spectrum for $gN \gtrsim 1$, there appears a 
Goldstone mode $A'$ (the translation mode of the soliton) 
associated with the symmetry breaking of the ground state, 
the breathing mode $B'$, and the second harmonic of the breathing mode $C'$.
In Fig.~\ref{fig1} (a) for $gN\gtrsim 1$, in contrast, a number of quasi-degenerate levels 
appear with the density of states peaking around the Bogoliubov levels; 
we denote the corresponding groups as $A, B$, and $C$, respectively. 
The basis states for the diagonalization are restricted to 
the angular-momentum states $l=0,\pm 1$ ($l_{\rm c}=1$) unless otherwise stated, 
and the field operator is given by 
${\hat \psi}(\theta)=({\hat c}_0+e^{i\theta}{\hat c}_1+e^{-i\theta}{\hat c}_{-1})/\sqrt{2\pi}$, 
where ${\hat c}_l$ is the annihilation operator of a boson with angular momentum 
$\hbar l$. 
The validity of this cutoff has been confirmed by the inclusion of 
higher angular-momentum states as shown in the inset of 
Fig.~\ref{fig1} (a) with $l_{\rm c}=2$~\cite{NP}, where the energy landscape and 
the degree of degeneracy are unchanged from those of $l_{\rm c}=1$. 


We denote the eigenstates of the Hamiltonian as $|{\cal L}\rangle_{\sigma}$, 
where ${\cal L}$ is the total angular momentum. 
The index $\sigma=A,B,\cdots$ labels the eigenstate in the ascending order of energy 
($E_{|{\cal L}\rangle_A} < E_{|{\cal L}\rangle_B} < \cdots$) for each ${\cal L}$.
The states $|{\cal L}\rangle_{\sigma}$ and $|-{\cal L}\rangle_{\sigma}$ 
are degenerate in the absence of a rotating drive. 
At $gN=0$, the eigenstates are the Fock states 
$|{\cal L}\rangle_{\sigma}=|n_1,n_{-1}\rangle$, where $n_{\pm 1}$ denote the 
numbers of atoms with angular momenta $l=\pm 1$ and ${\cal L}=n_1-n_{-1}$. 
The energies of the states $|n_1,n_{-1}\rangle$ are given by $n_1+n_{-1}\equiv j$, 
and these states are thus $(j+1)$-fold degenerate. 
The index $\sigma=A,B, \cdots$ corresponds to the number of $l=\pm 1$ pairs 
being given by  $(j-|{\cal L}|)/2=0,1,\cdots$. 
For $0< gN \lesssim 1$, the energy branches are characterized by $j$, 
and the $(j+1)$-fold degeneracy is almost maintained. 
As $gN$ approaches 1, each branch begins to ramify, and 
the energy landscape for $gN \gtrsim 1$ is characterized 
by the index $\sigma$ and ${\cal L}$ as 
$E_{|0\rangle_{\sigma}}\lesssim E_{|\pm 1\rangle_{\sigma}} 
\lesssim E_{|\pm 2\rangle_{\sigma}}\lesssim\cdots$. 
There is no Goldstone mode because the ground state possesses the translation symmetry, 
and the lowest excited states $|\pm 1\rangle_A$ have a finite energy gap 
$\Delta E\equiv E_{|\pm 1\rangle_A}-E_{|0\rangle_A}$, since the system is finite. 
However, the density of states above the ground state becomes higher for larger $N$, 
and the gap $\Delta E$ collapses as $1/N$ [Fig.~\ref{fig1} (c)].
The ground state is therefore unstable against excitations of the 
quasi-degenerate low-lying states. 


We construct the many-body counterparts of the bright soliton $|\Psi_{\theta}\rangle$ and 
the Goldstone mode $|\Phi_{\theta}\rangle$, such that $\langle\Psi_{\theta}|\Phi_{\theta}\rangle=0$ 
by superpositions of the ground and quasi-degenerate states: 
\begin{eqnarray}
|\Psi_{\theta}\rangle \!\!\!&\equiv&\!\!\! e^{-i{\hat L}\theta}
\left[\beta_0|0\rangle_A+\sum_{{\cal L}>0}
\beta_{\cal L}\left( |{\cal L}\rangle_A + |-{\cal L}\rangle_A\right)\right],\label{soliton}\\
|\Phi_{\theta} \rangle \!\!\!&\equiv&\!\!\! \frac{d}{d\theta}|\Psi_{\theta}\rangle
\!\!=\!\!-ie^{-i{\hat L}\theta}\sum_{{\cal L}>0}{\cal L}
\beta_{\cal L}\left(|{\cal L}\rangle_A-|-{\cal L}\rangle_A\right)\label{Goldstone}, 
\end{eqnarray}
where $\hat{L}\equiv \int d\theta \hat{\psi}^{\dagger}(\theta)(-i\partial_{\theta})\hat{\psi}(\theta)$ 
is the angular momentum operator, and $\beta_{\cal L}$'s satisfy $\sum_{\cal L} |\beta_{\cal L}|^2=1$. 
The energy cost associated with the superposition (\ref{soliton}), 
$E_{|\Psi_{\theta}\rangle}-E_{|0\rangle_A}
=\sum_{{\cal L}}|\beta_{\cal L}|^2\!
\left(E_{|{\cal L}\rangle_A}-E_{|0\rangle_A}\right)$, 
is on the order of $1/N$. 
This indicates that the symmetry breaking from the exact ground state $|0\rangle_A$ 
to the bright soliton $|\Psi_{\theta}\rangle$ costs little energy, 
and the superposition thus occurs by an ``infinitesimal'' perturbation 
on the order of $1/N$. 



\begin{figure}
\includegraphics[scale=0.45]{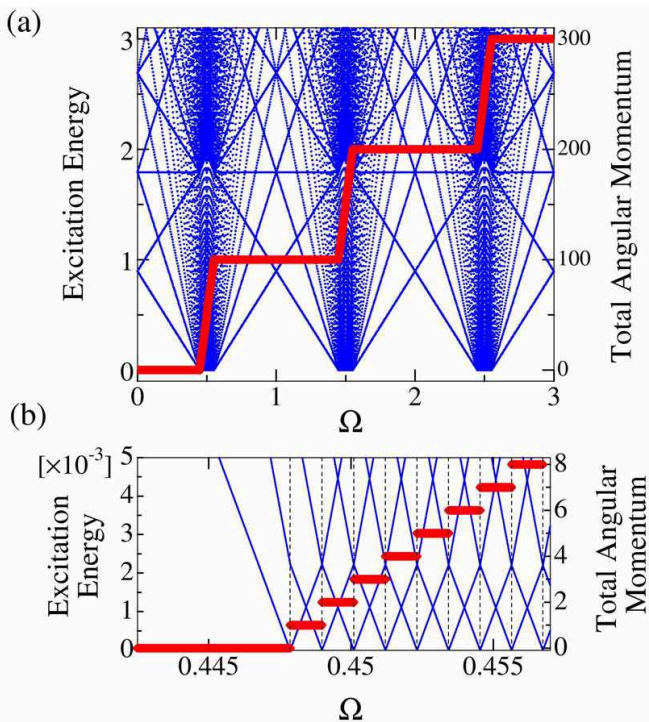}
\caption{
Low-lying spectrum (thin lines, left scale) and expectation value of 
the total angular momentum (thick line, right scale) of the ground state
versus the angular frequency of the rotating drive 
for $g=2\times 10^{-3}$ and $N=100$. 
(b) is an enlargement of (a). 
}
\label{fig2}
\end{figure}

The emergence of quasi-degenerate levels also plays a crucial role in 
breaking the quantized circulation. 
In the presence of a rotating drive with angular frequency $2\Omega$, 
the many-body ground state is either a single or 
fragmented condensate, depending on whether 
$f(g,N,\Omega)\equiv (1-gN)/2-2(\Omega-[\Omega+1/2])^2$ is positive or negative~\cite{rotation}, 
where $[\Omega+1/2]$ denotes the maximum integer that does not exceed $\Omega+1/2$. 
Figure~\ref{fig2} shows the total angular momentum 
of the ground state ${\cal L}_{\rm g}$ and 
low-lying eigenvalues of the Hamiltonian ${\hat H}-2\Omega\hat{L}+\Omega^2$ 
in the rotating frame. 
There clearly appear two distinct regimes where the density of states of excitations is 
low ($f>0$) and high ($f<0$) in the spectrum, and 
the circulation $h{\cal L}_g/m$ is quantized only when $f>0$. 
When the density of states above the ground state is sparse, 
the ground state cannot make a transition to states with higher angular momenta 
even if $\Omega$ is increased, 
since the term $-2\Omega{\hat L}$ is not large enough to make up for the excitation energy 
due to a large energy gap. 
The total angular momentum ${\cal L}_{\rm g}$ therefore 
does not increase with $\Omega$, and is quantized at integral multiples of $N$ 
for $f>0$. 
However, a number of branches with higher angular momenta 
decrease in energy as $f$ becomes negative; then, as $\Omega$ is increased, 
one branch after another takes the place of the ground level 
upon intersection [Fig.~\ref{fig2} (b)]. 
The angular momentum of the ground state ${\cal L}_{\rm g}$ thus 
increases stepwise each time the excitation energy collapses. 
The interval between collapses of the excitation energy, i.e., 
the width of the steps of ${\cal L}_{\rm g}$, 
becomes narrower as $N$ becomes larger, eventually resulting in the breaking of 
the quantized circulation, as indicated by the slopes in the thick line in Fig.~\ref{fig2} (a). 
The emergence of quasi-degenerate levels also induces symmetry breaking 
in a manner similar to Eq.~(\ref{soliton}). 
In fact, the ground state in the GP theory is a localized soliton for 
$f<0$~\cite{rotation}. 

\begin{figure}
\includegraphics[scale=0.45]{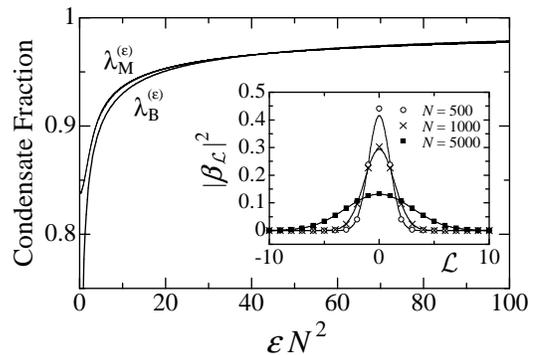}
\caption{
The largest eigenvalue of the reduced single-particle 
density matrix obtained by the exact diagonalization 
($\lambda_{\rm M}^{(\varepsilon)}$) and 
Bogoliubov theory ($\lambda_{\rm B}^{(\varepsilon)}$) 
versus $\varepsilon N^2$ for $gN=1.4$. 
Inset: Distribution of 
$|\beta_{\cal L}|^2=|{}_A\langle {\cal L} | \Psi(\varepsilon)\rangle|^2$ 
with $\varepsilon=1\times 10^{-4}$. 
The solid curves depict Eq.~(\ref{Gaussian}) with Eq.~(\ref{widthP}).
}
\label{fig3}
\end{figure}

We next investigate the effect of a symmetry-breaking perturbation 
by diagonalizing the Hamiltonian $\hat{H}+\varepsilon \hat{V}$ where 
$\hat{H}$ is given by Eq.~(\ref{hamiltonian}) and 
$\hat{V}=\int d\theta \hat{\psi}^{\dagger}(\theta)\cos\theta\hat{\psi}(\theta)$. 
As the perturbation is switched on, the ground state begins to localize by 
superposing the quasi-degenerate states. 
The largest eigenvalue $\lambda_{\rm M}^{\rm (\varepsilon)}$ 
of the single-particle density matrix is plotted as a function of $\varepsilon N^2$ 
in Fig.~\ref{fig3}. 
The fragmented condensate approaches the single condensate as $\varepsilon N^2$ increases. 
We note that $\lambda_{\rm M}^{(\varepsilon)}$ depends only on $\varepsilon N^2$, which 
indicates that a single condensate is realized for $\varepsilon 
\gtrsim N^{-2}$. The fragmented condensate with $N\gg 1$ is hence 
very fragile against the symmetry-breaking perturbation.


The condensate fraction $\lambda_{\rm M}^{(\varepsilon)}$ and the distribution of 
$\beta_{\cal L}$ in Eq.~(\ref{soliton}) can be derived analytically with 
the Bogoliubov approximation~\cite{derivation}.
Since the frequency of the translation mode of the broken-symmetry soliton is 
proportional to $\varepsilon^{1/2}$ and small, 
the number of excited atoms in this mode is much larger than 
in the other modes. We thus consider only the translation mode, and 
obtain the depletion as 
$(2N)^{-1} \varepsilon^{-1/2} [(2F)^{-1/2} - \varepsilon^{1/2} + 
O(\varepsilon)]\equiv 1-\lambda_{\rm B}^{(\varepsilon)}$, where 
$F \equiv 7 (3gN+4)^{1/2} / [(5gN+2)(2gN-2)^{1/2}]$
is a decreasing function for $gN > 1$. 
The behavior of $\lambda_{\rm B}^{(\varepsilon)}$ is in excellent agreement 
with the numerical result as shown in Fig.~\ref{fig3}.
Assuming that $\beta_{\cal L} |{\cal L}\rangle_A$ 
in Eq.~(\ref{soliton}) corresponds to the projection of the Bogoliubov 
ground state on the angular momentum ${\cal L}$, i.e., 
$\beta_{\cal L}|{\cal L}\rangle_A \simeq\int\frac{d\theta}{2\pi} e^{i ({\cal L} - \hat{L}) \theta} |\Psi_{\rm B}\rangle$,
we obtain the distribution of $\beta_{\cal L}$ as
\begin{eqnarray}
|\beta_{\cal L}|^2\simeq\frac{1}{\sqrt{2 \pi d^2}}\exp\left[-\frac{{\cal L}^2}{2 d^2}\right],
\qquad\qquad\qquad\label{Gaussian}\\
d^2=N \varepsilon^{1/2} \frac{2(gN-1)}{7gN} 
\left[(2F)^{1/2} + F \varepsilon^{1/2} + O(\varepsilon)\right]\label{widthP}, 
\end{eqnarray}
which is in excellent agreement with the numerical results as shown in the inset of Fig.~\ref{fig3}. 
The behavior of the width $d\propto N^{1/2} \varepsilon^{1/4}$ 
indicates that the 
center-of-mass fluctuation is proportional to $N^{-1/2} \varepsilon^{-1/4}$.

\begin{figure}
\includegraphics[scale=0.5]{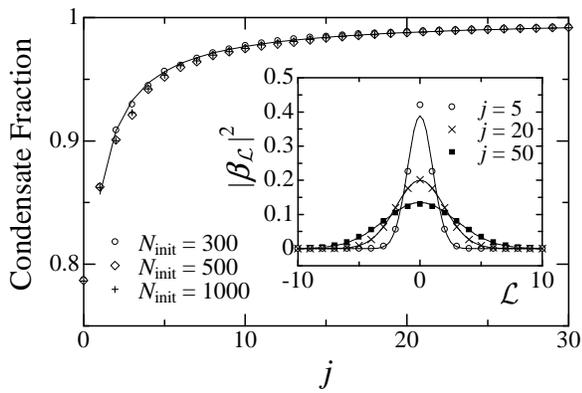}
\caption{
Ensemble-averaged largest eigenvalue $\bar{\lambda}_{\rm M}^{(j)}$ 
of the reduced single-particle density matrix 
versus the number of quantum measurements $j$ for $gN_{\rm init}=1.6$. 
The solid curve denotes the analytical result (\ref{anaM}). 
Inset: Ensemble-averaged value of 
$|\beta_{\cal L}|^2=|{}_A\langle {\cal L} | \Psi^{(j)}\rangle|^2$ 
after $j=5,20$ and $50$-time measurements. 
The solid curves depict Eq.~(\ref{Gaussian}) with 
$d^2 = 4 j (gN_{\rm init}-1)/(7gN_{\rm init})$.
}
\label{fig4}
\end{figure}


Finally, we investigate what happens to the exact ground state 
under the action of a quantum measurement~\cite{PD}. 
The ground state $|0\rangle_A$ 
is prepared as an initial state with the number of atoms $N_{\rm init}$ 
so that $gN_{\rm init} > 1$. 
Suppose that one atom is detected at position $\theta_j$ in the $j$-th measurement. 
The postmeasurement state $|\Psi^{(j)} \rangle$ is
related to the premeasurement one $|\Psi^{(j-1)} \rangle$ by 
$|\Psi^{(j)}\rangle = 
\hat{\psi}(\theta _j)|\Psi^{(j-1)}\rangle/\sqrt{\langle \Psi^{(j-1)}|\hat{\psi}^{\dagger}(\theta_j)\hat{\psi}(\theta_j)|\Psi^{(j-1)}\rangle}$. 
The normalized single-particle density is given by 
$n^{(j)}(\theta)=\langle\Psi^{(j)}|\hat{\psi}^{\dagger}(\theta)
\hat{\psi}(\theta)|\Psi^{(j)}\rangle/N_j$, 
and the position of the $(j+1)$-th measurement $\theta_{j+1}$ is probabilistically determined 
after the $j$-th measurement according to the probability distribution
$n^{(j)}(\theta)$. 
Since each run is a stochastic process, 
we numerically perform sequential runs of independent simulations 
and take the ensemble average for each $j$. 
We find that the ensemble-averaged value of the condensate fraction 
$\bar{\lambda}_{\rm M}^{(j)}$ is independent 
of $N_{\rm init}$ for a fixed $gN_{\rm init}$, and that it monotonically 
increases as shown in Fig.~\ref{fig4}. 
Therefore, when $N_{\rm init}\gg 1$, $\bar{\lambda}_{\rm M}^{(j)}$ reaches 
the order of one for $j/N_{\rm init} \ll 1$, indicating that 
the fragmented condensate rapidly becomes a single condensate by the action of quantum measurements. 
The single-particle density $n^{(j)}(\theta)$ localizes and 
the state $|\Psi_{\theta}^{(j)}\rangle$ reaches the form of Eq.~(\ref{soliton}). 


Suppose that the initial state is a uniform superposition of the soliton state 
$\int d \theta_0 A(\theta_0) \phi_{\rm sol}^{\rm GP}(\theta-\theta_0)$, 
where $\theta_0$ is the center of mass and $A(\theta_0)$ is a constant for $j=0$. 
We can show~\cite{derivation} that the probability distribution is changed by 
the $j$-time measurements as 
$A^2_j(\theta_0) \propto \exp (-4 c j \theta_0^2)$ with $c\equiv 2(gN_{\rm init}-1)/(7gN_{\rm init})$, 
and hence the distribution of $|\beta_{\cal L}|^2$ reduces to 
the Gaussian form~(\ref{Gaussian}) with $d^2 = 2 c j$, 
i.e., $d\propto \sqrt{j}$ for $j \lesssim \sqrt{N_{\rm init}}$. 
The center-of-mass fluctuation thus reduces to 
$\Delta \theta_0 \propto j^{-1/2}$ by the $j$-time measurements.
The condensate fraction can also be derived analytically~\cite{derivation} as 
\begin{eqnarray}\label{anaM}
\lambda_{\rm M}^{(j)}&=&\frac{1}{2}
\left[1-c+c e^{-2\alpha}\right.\nonumber\\
&+&\!\!\!\!\!\left.\sqrt{(1-c+c e^{-2\alpha})^2-4c(1-2c)(1-e^{-\alpha})^2}\right], 
\end{eqnarray}
where $\alpha=1/(8cj)$. 
These results are in excellent agreement with the numerical ones as shown in Fig.~\ref{fig4}. 



In conclusion, we employed the exact diagonalization method 
to investigate the simultaneous emergence of 
a bright soliton and the Goldstone mode in a 1D attractive BEC. 
We found that the existence of a number of quasi-degenerate states 
above a critical strength of attractive interaction makes 
the ground state fragile against symmetry-breaking perturbations. 
By introducing a symmetry-breaking perturbation or quantum measurements, 
we showed that a localized state with an enhanced condensate fraction 
is generated by a superposition of quasi-degenerate many-body states, 
and that the resulting state closely resembles the Bogoliubov ground state. 
The mechanism underlying a partial breaking of the quantized circulation 
is also elucidated in terms of the many-body spectrum. 



This work was supported by a 21st Century COE program at
Tokyo Tech ``Nanometer-Scale Quantum Physics'', 
Special Coordination Funds for Promoting Science and Technology, and 
a Grant-in-Aid for Scientific Research (Grant No. 15340129) by the
Ministry of Education, Culture, Sports, Science and Technology.




\begin{thebibliography}{30}

\bibitem{NS}
P.~Nozi\`{e}res and D.~Saint James, J.~Phys. (France) {\bf 43}, 1133 (1982). 

\bibitem{fr}
N.K.~Wilkin, J.M.F.~Gunn, and R.A.~Smith, Phys. Rev. Lett. {\bf 80}, 2265 (1998);
D.S.~Rokhsar, e-print cond-mat/9812260; 
C.K.~Law {\it et al}., Phys. Rev. Lett. {\bf 81}, 5257 (1998); 
E.J.~Mueller, T.L.~Ho, G.~Baym, and M.~Ueda, unpublished; 
M.~Koashi and M.~Ueda, Phys. Rev. Lett. {\bf 84}, 1066 (2000); 
T.-L.~Ho and S.-K.~Yip, Phys. Rev. Lett. {\bf 84}, 4031 (2000); 
S.~Ashhab and A.J.~Leggett, Phys. Rev. A {\bf 68}, 063612 (2003). 

\bibitem{UL}
M.~Ueda and A.J.~Leggett, Phys. Rev. Lett. {\bf 83}, 1489 (1999).

\bibitem{PO} 
O.~Penrose and L.~Onsager, 
Phys. Rev. {\bf 104}, 576 (1956).

\bibitem{BS-EX}
 K.~E.~Strecker, G.~B.~Partridge, A.~G.~Truscott, 
and R.~G.~Hulet, Nature (London) {\bf 417}, 150 (2002); 
 L.~Khaykovich, F.~Schreck, G.~Ferrari, T.~Bourdel, 
J.~Cubizolles, L.~D.~Carr, Y.~Castin, C.~Salomon, 
Science {\bf 296}, 1290 (2002).

\bibitem{BS-MF}
U.~Al~Khawaja, H.T.C.~Stoof, R.G.~Hulet, 
K.E.~Strecker, and G.B.~Partridge, 
Phys. Rev. Lett. {\bf 89}, 200404 (2002); 
L.D.~Carr and J.~Brand, 
Phys. Rev. Lett. {\bf 92}, 040401 (2004). 
V.E.~Zakharov and A.B.~Shabat, 
Zh. Eksp. Teor. Fiz. {\bf 61}, 118 (1971)
[Sov. Phys. JETP {\bf 34}, 62 (1972)].

\bibitem{symmetry}
B.~Bernu, C.~Lhuillier, and L.~Pierre, Phys. Rev. Lett. {\bf 69}, 17 (1992); 
E.H.~Rezayi, F.D.M.~Haldane, and K.~Yang, Phys. Rev. Lett. {\bf 83}, 1219 (1999); 
Y.~Castin, e-print cond-mat/0012040; 
N.R.~Cooper, N.K.~Wilkin, and J.M.F.~Gunn, Phys. Rev. Lett. {\bf 87}, 120405 (2001); 
T.~Nakajima and M.~Ueda, Phys. Rev. Lett. {\bf 91}, 140401 (2003). 

\bibitem{QPT} 
R.~Kanamoto, H.~Saito, and M.~Ueda, 
Phys. Rev. A {\bf 67}, 013608 (2003).

\bibitem{NP}
We employed the Lanczos algorithm in TITPACK Ver. 2 
developed by H.~Nishimori. 

\bibitem{rotation} 
R.~Kanamoto, H.~Saito, and M.~Ueda, 
Phys. Rev. A {\bf 68}, 043619 (2003); 
P.F.~Kartsev, Phys. Rev. A {\bf 68}, 063613 (2003). 

\bibitem{derivation}
Details of the derivations will be reported elsewhere.

\bibitem{PD}
J.~Javanainen and S.M.~Yoo, 
Phys. Rev. Lett. {\bf 76}, 161 (1996); 
Y.~Castin and J.~Dalibard, 
Phys. Rev. A {\bf 55}, 4330 (1997). 

\end{thebibliography}
\end{document}